# Paste rheology and surface charge of calcined kaolinite

Y. Demeusy[*], S. Gauffinet, and C. Labbez

*ICB UMR 6303 CNRS, Université de Bourgogne, FR-21000 Dijon, France*
*Email: yannick.demeusy@u-bourgogne.fr*

**ABSTRACT**

Partial substitution of the clinker in the cement by a supplementary cementitious material (SCM) is one of the main solutions to reduce the carbon footprint. Calcined kaolinite is a good candidate due to its availability and relatively high reactivity compared to other SCMs. The main issue with these calcined clay types of cements is the high-water demand at low clinker factors, a problem which remains not well understood. In this proceeding, we will show the role played by electrostatic interactions in the paste stiffening using as a model system pure calcined kaolinite paste prepared at various pH as well as salt types and concentrations. The study combines dynamic rheometry measurements in strain-sweep modes, surface charge characterization using potentiometric titration and electrophoretic measurements as well as calculations of inter-particle interactions using Monte-Carlo (MC) simulations in the framework of the primitive model. The calcined kaolinite is found to possess a negative permanent charge, presumably due to the Si(IV)/Al(III) substitution, and a titratable charge (as due to ionization of silanol and aluminol surface groups) with a point of zero proton charge at pH 4.65. In conditions relevant for cement paste, the calcined clay bear a strong negative charge ~ 300 mC/m$^2$. The rheological measurements reveal that the paste stiffening is highly dependent on the pH, salt concentration and type as expected for systems controlled by electrostatic interactions. The stiffness increases with the salt concentration at natural pH and is the largest in solutions buffered with $Ca(OH)_2$, that is at high $Ca^{2+}$ concentrations and pH where the negative charge of the calcined clay is the strongest. The MC simulations of the inter-particle interactions are found to qualitatively explain the observed variation in the paste stiffness.

**KEYWORDS:** *calcined clay, kaolinite, surface charge, rheology*

**1. Introduction**

The major source of $CO_2$ during cement manufacture is the decomposition of limestone into calcium oxide, $CaCO_3 \rightarrow CaO + CO_2$ (Flower et al 2007). This means that the emission of nearly 0.8 ton of $CO_2$ per ton of produced cement clinker are intrinsic to the material and cannot be reduced by the modernization of the plants, i.e., change of energy source. This can be achieved instead by substituting part of the cement clinker by supplementary cementitious materials (SCM) and, among those, one of the most promising is calcined clay (Scrivener et al (2018). Calcined clay cements are already produced and in use in a few major European and world countries. However, above 30% of clinker substitution serious implementation problems are encountered mainly due to the high water demand of calcined clays. Numerous hypotheses have been formulated to explain this high water demand but none of them gives full satisfaction or has been clearly demonstrated. In this study we propose to investigate the link between the interfacial properties of calcined clays and in particular their surface electric charge with their rheological properties in paste. To do this we combined dynamic rheometry measurements in strain-sweep modes, surface charge characterization using potentiometric titration and electrophoretic measurements as well as calculations of inter-particle interactions using Monte-Carlo (MC) simulations. All these measurements and simulations were conducted with pure calcined kaolinite in various aqueous solutions where the pH, salt type and concentration were varied systematically.

**2. Methodology**

## 2.1 Clay and preparation

The material used in this study is a commercial kaolinite with a purity of 96% (Sigma-Aldrich 03584), calcined at 700°C for 2h and air quenched. The full dehydroxylation of the clay was controlled by TGA (not shown). The clay was cleaned by dispersing and mixing it in a solution of 10 mM HCl for 10 min. The clay sample was then vacuum filtered and rinsed with a solution of 1M NaCl. This procedure was repeated three times. Finally, the washed clay sample was purified by dialysis for 3 weeks in deionized water renewed every day. The composition of the clay as obtained by X-Ray fluorescence (XRF) is given in Table 1.

Table 1 Composition of kaolinite obtained by XRF

| Oxyde | $SiO_2$ | $Al_2O_3$ | $K_2O$ | $Fe_2O_3$ | $P_2O_5$ | $TiO_2$ | PbO | BaO | MgO | CaO |
|---|---|---|---|---|---|---|---|---|---|---|
| Mass Concentration | 53.56% | 41.64% | 2.05% | 0.66% | 0.57% | 0.52% | 0.37% | 0.22% | 0.15% | 0.12% |

## 2.2 Rheometry

Dynamic rheometry in the strain sweep mode was used to measure the young modulus of the calcined kaolinite pastes using a parallel plate geometry. A sample consists of a disk of paste with a diameter of 25 mm and a 2 mm gap. Given these dimensions and the characteristics of the rheometer, the lowest measurable strain was 0.01% ($10^{-4}$). The liquid to solid ratio of the clay paste was set to L/S = 1.33. The rheological measurements were made with a controlled strain rheometer (ARES-G2 TA instrument) whereby the sample is submitted to a sinusoidal strain with an oscillatory frequency of $\omega = 10$ rad.s$^{-1}$, and the stress response is recorded. The presented results are an average of at least 3 independent measurements.

## 2.3 Potentiometric titration

The titratable surface charge density was measured by potentiometric titration of 1g calcined kaolinite dispersed in 50mL aqueous solution containing NaCl as a background salt. Titration was performed using as titrant, 10 mM NaOH and 10 mM HCl solutions containing the same NaCl background salt concentration as the calcined kaolinite dispersion contained in the reactor. The titration was measured at background salt concentration of 1, 10 and 100 mM.

## 2.4 Electrophoretic mobility

The electrophoretic mobility was measured by dynamic light scattering using a Malvern zetasizer (nano zs) on diluted suspensions of well dispersed calcined clay particles. The latter was obtained by dispersing 2g of clay in 250mL of deionized water which was then decanted for 1h. The 3 upper centimeters were used for the zeta potential analysis.

## 2.5 Inter-particular forces simulations

The simulations within the so-called slit model which consists in two infinite parallel surfaces (the clay surfaces) separated with an electrolyte solution in equilibrium with a reservoir of set concentration. The model was solved with Monte-Carlo simulations in the grand canonical ensemble (Valleau et al 1980) with use of the Metropolis algorithm (Metropolis et al 1953). The calculated double layer osmotic pressure is converted to a double layer force by applying the Dejarguin approximation (Russel 1989). The total interparticle force is obtained by summing the latter and the van der Waals force. Further details on the model and simulations can be found elsewhere (Jönsson et al 2005). In the simulation an Hamaker constant equal to that of silica was used, $H = 2.4 \times 10^{-21}$ J.

## 3. Results & Discussion

Fig. 1-a shows the typical nonlinear oscillatory response (strain sweep measurements) of the calcined clay pastes in Ca(OH)$_2$ buffered solution. The clay paste has a viscoelastic behavior similar to that found with C-S-H gels. At low strain, the Young modulus (G') and the loss (viscous) modulus (G''), show a typical plateau characteristic of the linear viscoelastic region (LVR). G' is further found larger than G'' (G'>G''). In this region the paste thus behaves as an elastic solid where deformations are reversible, i.e., the microstructure of the paste remains unchanged. As the strain increases outside the LVR, above the critical strain, the microstructure of the paste is changed and a drop in G' is observed but G' remains larger than G". In this region the paste behaves as a plastic solid where deformations are irreversible. As the strain is increased further, G" crosses G' and becomes larger than G'. At this crossing point the paste starts to flow/yield and the yield stress can be estimated, see Fig. 2-b. The measured young modulus as a function of the strain in strain sweep mode of calcined clay pastes prepared in various electrolyte solutions are shown in Fig. 1-b in comparison to the clay paste in pure water at ω = 10 rad/s. The lowest values of G' are obtained for pure water and are observed, for the same pH (pH~4.5), to increase with the ionic strength, i.e. G'(100 mM $Na_2SO_4$) < G'(1 M $NaNO_3$). The highest G' values are observed for the paste buffered with $Ca(OH)_2$, that is for high pH and calcium concentration (pH = 12.7 and $Ca^{2+}$ = 22 mM). In these conditions G' is 2 orders of magnitude higher than in pure water and 1 order of magnitude higher than in concentrated salt solution at pH ~ 4.5. These G' variations with the salt concentration and pH indicate that the electrostatic interactions play an important contribution to the change in the microstructure and bound strength between the clay particles in the paste.

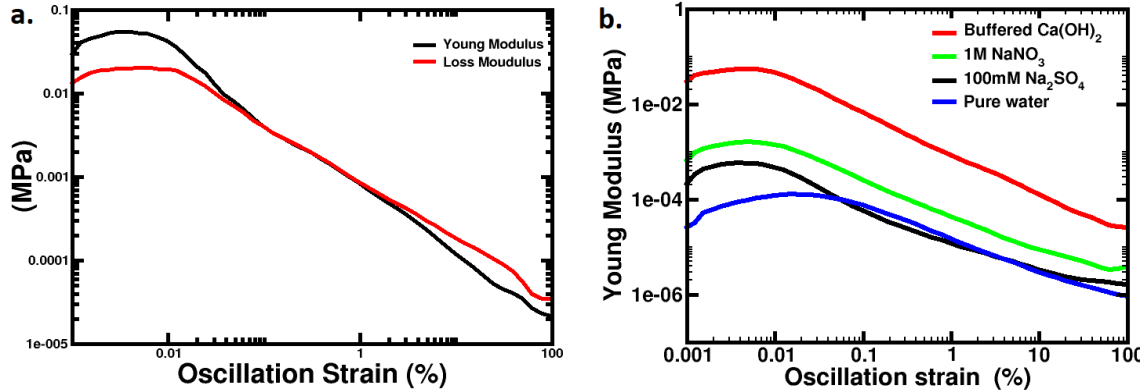

Figure 1 a- Strain-sweep measurements of the young modulus (G') and loss modulus (G") of a calcined Strain-sweep measurements of the calcined kaolinite paste in the presence of $Ca(OH)_2$ solution. b- Young modulus versus strain of calcined kaolinite pastes in the presence of various electrolytes and in pure water with a liquid/solid ration of 1.33. The pH of the paste in pure water, 100 mM $Na_2SO_4$ and 1M $NaNO_3$ is the same, pH ~4.5 while it is pH = 12.7 in buffered $Ca(OH)_2$ conditions.

The charging behavior of the calcined clay at various pH and background salt concentration was thus studied by combining potentiometric titration and electrophoretic mobility. The potentiometric titration data shown in Fig. 2-a confirms that the calcined clay bear a pH dependent surface charge. The titratable surface charge ($\sigma_{tit}$) is further found to be amphoteric with a neutral titratable charge, the point of zero net proton consumption (PZNPC), between pH 4.5 and 5. At pH values below the PZNPC $s_{tit}$ is positive and negative above the PZNPC. and becomes strongly negative at high pH values, that is in the cement pore solution conditions. The PZNPC is further found to decrease with the ionic strength but the variation of the electric charge with the same remains weak. The shift of the PZNPC and the small change of $\sigma_{tit}$ with the ionic strength indicate that in addition to a titratable surface charge the calcined clay particles bear a permanent structural charge ($\sigma_{st}$). The decrease of the PZNPC with the ionic strength further indicates that $\sigma_{st}$ is of negative, presumably due to the substitution of silicon by aluminum. The value of isoelectric point (IEP), where the diffuse potential (zeta potential) and the total charge of the clay particles are null ($\sigma_{tit} + \sigma_{st} = 0$), obtained by electrophoretic mobility confirms the existence and sign of $s_{tit}$. As shown in Fig. 2-a the IEP is found at a pH value, pH ~ 2.8, much below that of the PZNPC. Indeed, the negative sign of the zeta potential at the pH value of the PZNPC can only be explained by $\sigma_{st} < 0$.

The interparticle forces between the clay particles were calculated, not shown. In pure water, the mean force is found to be purely repulsive and long range due to the long range electrostatic repulsion. When the ionic strength I is increased, I(0.1M $Na_2SO_4$) < I(1M $NaNO_3$), the electrostatic interactions are screened and the force becomes attractive due to the short range van der Waals interactions. In buffered $Ca(OH)_2$ solution (pH = 12.7 and $Ca^{2+}$ = 22 mM) the force becomes strongly attractive due to the appearance of attractive ion-ion correlation forces. The yield stress and young modulus can be correlated to, respectively, the contact mean force and curvature of the interaction-free-energy well between the particles. A rather good correlation between the measured yield stresses and simulated contact forces is obtained, see Fig. 2-a. The same correlation is found for G' (not shown). The rheological behavior of the calcined clay pastes can thus be explained with the simple electrostatic model used. In pure water the interaction between the calcined clay particles are purely repulsive due to strong electrostatic repulsions which "lubricate" the contacts between the particles. The yield stress and G' are then the lowest. At neutral pH, when a salt is added to the solution the electrostatic repulsions are screened and a percolated network of clay particles is formed due to the prevalence of the attractive van der Waals forces over the repulsive electrostatic forces. The bounds between the particles strengthen and thus the yield stress and G' increase with increasing the ionic strength, e.g., G'(100 mM $Na_2SO_4$) < G'(1M $NaNO_3$). When the pH is increased the surface charge density of the calcined clay increases in absolute value. In these conditions when a calcium salt is added, strong bounds between the particles are formed due to strong attractive ion-ion correlation forces. This explains the highest values of yield stress and G' observed in the solution buffered with $Ca(OH)_2$.

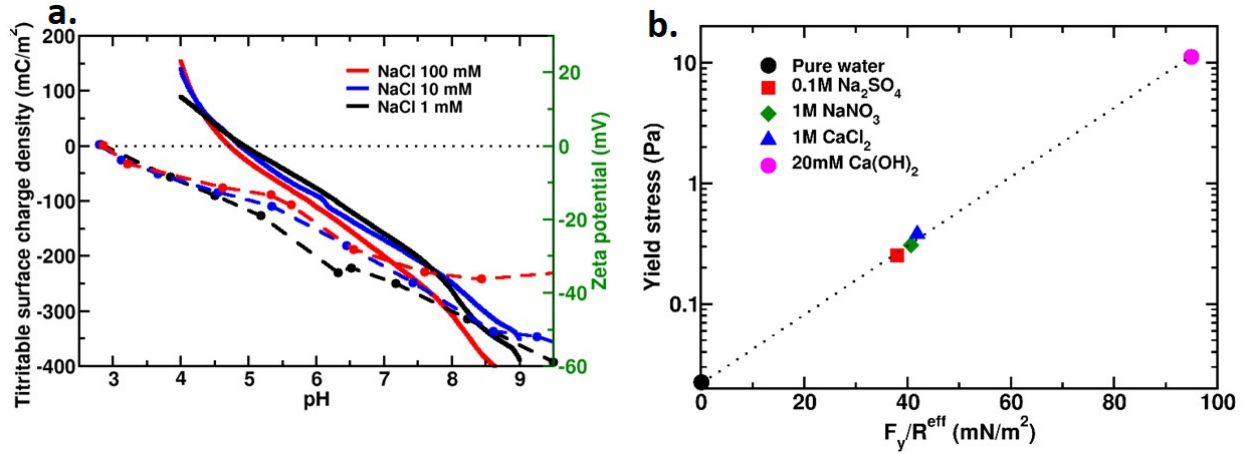

Figure 2 (a.) Titratable surface charge density and zeta potential of the calcined kaolinite in a 1/10/100mM NaCl background salt solutions as obtained by potentiometric titration and electrophoretic mobility. Plain curves: titratable surface charge density; Dashed curves: zeta potential. (b.) Measured yield stress versus simulated contact force. In the case of pure water, where the interactions are purely repulsive, the difference of the forces at separations of 1 micron and 500 nm is taken. These positions corresponds to the equilibrium position of a particle surrounded by 8 neighbors and the minimal distance for the same to escape such a "cage".

## 5. Conclusions

Calcined kaolinite are found to bear a titratable surface charge, which is positive at low pH and strongly negative at high pH, as well as a permanent structural surface charge of negative sign. The titratable charge comes from the ionization of silanol (Si-OH = Si-O$^-$ + H$^+$) and aluminol (Al-OH$^{1/2+}$ = Al-O$^{1/2-}$ + H$^+$) surface groups. The structural charge comes presumably from the substitution of silicon by aluminum. One can expect that the latter varies with the origin of the clay mineral. The high water demand of the calcined clay paste can be rationalized by the formation of a percolated network. In particular, the increase of the yield stress and stiffness of the paste with ionic strength, pH and calcium concentration, is well capture with a simple model involving the superposition of electrostatic and van der Waals interactions between two clay particles. The yield stress and young modulus of the paste are observed to be the highest in solutions buffered with $Ca(OH)_2$ where the charge of the particles and the attractive ion-ion correlation forces are found to be the strongest.